\documentstyle[epsfig,aps,preprint,prl]{revtex}
 
\begin{document}

\title{Coupling of Dipole Mode to $\gamma$-unstable Quadrupole Oscillations}

\author{C.E. Alonso$^1$, M.V. Andr\'es$^1$, J.M. Arias$^1$, 
E.G. Lanza$^{1,2}$, and A. Vitturi$^3$}

\bigskip

\address{$^1$ Departamento de F\'{\i}sica At\'omica, Molecular y Nuclear,
Facultad de F\'{\i}sica \\ Universidad de Sevilla, Apartado~1065, 
41080 Sevilla, Spain\\
$^2$ INFN, Sezione di Catania, I-95129 Catania, Italy \\
$^3$ Dipartimento di Fisica and INFN, Padova, Italy}

\date{\today}
\maketitle

\begin{abstract}
The coupling of the high-lying dipole mode to the low-lying
quadrupole modes for the case of deformed $\gamma$-unstable nuclei is
studied. Results from the geometrical model are compared to those
obtained within the dipole boson model. Consistent results are
obtained in both models. The dipole
boson model is treated within the intrinsic frame, with subsequent  
projection onto the laboratory frame. As an application, calculations
of photonuclear cross-sections in $\gamma$-unstable nuclei are presented.
\end{abstract}

\vspace{3cm}

\noindent
{\bf PACS numbers: 21.60.Ev, 21.60.Fw, 21.10.Re}
 
\newpage

\section{Introduction}
Low-lying quadrupole oscillations in nuclei have been studied extensively
with all kinds of nuclear structure models. Among the more
popular are the geometrical model\cite{BMII} and the interacting boson
model (IBM)\cite{IA87}.
Although rather different in their microscopic interpretation and
formulation, both of them have been extremely successful in describing
low-lying energy levels associated to quadrupole oscillations.
In three different situations both models provide analytical results
that can be identified in the geometrical model as those corresponding
to spherical\cite{BMII}, well quadrupole  deformed \cite{BMII} and 
deformed $\gamma$-unstable shapes\cite{WJ56}. The corresponding
situations occur in IBM for the 
U(5)\cite{AIu5}, SU(3)\cite{AIsu3} and O(6)\cite{AIo6} limits
respectively. Connection between both models has been established by
exploring the geometrical content of the IBM by using coherent states
\cite{GK80,DSI80,BM80}.
 
High-lying dipole states are studied in the context of the
geometrical model by using the dynamic collective model within the
hydrodynamical approach\cite{DG64,DG64b}. In the IBM context these 
states can be
studied with the so called dipole boson model
\cite{MW82,SH83,Mai84,SH87}, which includes a dipole
$p$ boson into the usual IBM space (quadrupole $s$ and $d$ bosons).
The connection between both approaches, especially for the case of well
deformed nuclei, has already been discussed\cite{SH87}.

In this paper we are interested in studying the coupling of the dipole
mode to a $\gamma$-unstable quadrupole form. The results of the 
geometrical 
model will be presented and the study within the dipole boson model will
be performed both in the laboratory frame and in the intrinsic frame 
projecting afterwards onto the laboratory frame. 

The paper is structured as follows. In Section 2 the coupling of the
dipole mode to a $\gamma$-unstable rotor is analysed within the 
geometrical
model. In Section 3 the coupling of the dipole mode to an O(6)
quadrupole deformed nuclei  is
investigated in the interacting boson model.  This is
done in two steps, first we study the problem in the intrinsic frame
and then the projection to the laboratory frame is performed. Comparisons  of
exact results and those obtained in the intrinsic frame plus projection 
for different sets of parameters are presented and discussed. 
In Section 4 examples of calculations for photonuclear cross-section
are presented, and the different approximations compared. 
Finally, in Section 5 the paper is 
summarised and the main conclusions assessed.

\section{Coupling the dipole mode to a $\gamma$-unstable rotor in the 
geometrical model}

The interaction between dipole and quadrupole oscillations has been
studied in the geometrical model. The coupling to the quadrupole degree
of freedom splits the energy of the dipole mode, and within the linear coupling 
in the quadrupole amplitude\cite{BMII} the energies 
of the three resulting dipole resonances as seen from the intrinsic frame are
\begin{equation}
E_{1\mu}(\beta,\gamma)= \epsilon_1 
\left[1+ \left(\frac{2}{15}\right)^{1/2}~\frac{\kappa_1}{C_1} 
~\beta \cos \left(\gamma - \frac{2 (\mu+2)
\pi}{3}\right)\right]~~;
~~\mu=-1,0,1~~,
\end{equation} 
where $\epsilon_1$ is the unperturbed dipole energy, $C_1$ is the
restoring force parameter, and $\kappa_1$ is the dipole--quadrupole 
coupling coefficient. A plot of these energies with respect to the unperturbed
value $\epsilon_1$ is given in Fig. 1 as a function
of the asymmetry parameter $\gamma$, for a fixed value of the quadrupole
deformation parameter $\beta$

A detailed structure of the dipole absorption line can be obtained 
from a numerical diagonalization of the coupled dipole--quadrupole hamiltonian.
A simple model in the strong coupling approximation for extracting 
the dipole strength distribution from
the intrinsic values is provided by the procedure of 
averaging the three contributions over
the different quadrupole $\beta$ and $\gamma$ deformations, according
to the ground state distribution.  
The relative strength of the dipole 
transition probability $B(E1; 0 \rightarrow 1^-_n)$ 
per unit of energy is in this way proportional to
\begin{equation}
P(E) = \frac{1}{3} \sum_{\mu=-1}^1
\int \int \phi_0^2(\beta, \gamma) \delta\left(E-E_{1 \mu}(\beta,\gamma)\right)
\beta^4 d\beta |\sin 3\gamma| d\gamma~,
\label{BE1BM}
\end{equation}
where $\phi_0(\beta, \gamma)$ is the wave function of the ground state 
resulting from the quadrupole energy surface.
The specific case of $\gamma$-unstable rotor occurs when 
the intrinsic energy surface is isotropic in the $\gamma$ variable, while 
displaying a sharp minimum for deformation $\beta_0$. Assuming the extreme 
situation of a very sharp minimum, we can in first approximation keep fixed
the value of $\beta_0$ and only average over the $\gamma$ variable.
In this adiabatic picture 
one  obtains the probability distribution shown
in Fig. 2. 
The distribution is symmetric around $\epsilon_1$,  
with  three maxima, two at the edges (at energies 
$\epsilon_1 \pm \Delta$) and one in the center (at energy 
$\epsilon_1$), with
\begin{equation}
\Delta  =  \epsilon_1 \sqrt{2 \over 15} {\kappa_1 \over C_1} \beta_0 ~,
\end{equation}
and with zeros occurring at energies 
$\epsilon_1 \pm {\Delta \over 2}$.
The resulting total splitting $2\Delta$ is therefore directly proportional 
to the quadrupole deformation $\beta_0$, as in the case of a static quadrupole 
deformation. 
Note that although the probability at the edges is 3 times the one at
the center, the total probability is anyway 
exactly equally 
divided into the three bumps.  It may be worth noticing that,
aside from the broadening of the three peaks, the predicted situation 
in the $\gamma$-unstable case is actually
rather similar to the prediction associated with a static quadrupole deformed
system with $\gamma$ equal to 30 degrees. This similarity of the two situations
is a feature that also appears for other observables and has been discussed
elsewhere\cite{MGO94,BrOt97}.   
 
\section{Coupling the dipole mode to an O(6) quadrupole deformed
nucleus in IBM}

To include the dipole degrees of freedom in the usual IBM, one negative
parity dipole $p$--boson ($l=1$) has to be included in addition to the
positive parity quadrupole $s$-- ($l=0$) and $d$-- ($l=2$) bosons.
For a boson system with $N$ quadrupole bosons plus $1$ dipole
boson, the hamiltonian can be written as
\begin{equation}
H= H_{sd} + H_{p} + V_{sd-p}~,
\end{equation}
where $H_{sd}$ is the usual IBM hamiltonian for quadrupole degrees of
freedom, $H_{p}$ is the dipole hamiltonian given by
\begin{equation}
H_{p} = \epsilon_1 \hat n_p ~,
\end{equation}
and $V_{sd-p}$ is the
interaction between quadrupole and dipole degrees of freedom. The
operator $\hat n_p$ is the $p$-boson number operator, which in this
case is $0$ or $1$, and $\epsilon_1$ is the unperturbed dipole
energy. For the
interaction a simple quadrupole-quadrupole form is assumed
\begin{equation}
V_{sd-p}= -\kappa ~ Q^{(sd)} \cdot q^{(p)}~,
\end{equation}  
where $Q^{(sd)}$ is the usual IBM quadrupole operator
\begin{equation}
Q^{(sd)}_{\mu} = \left( s^\dagger \tilde d + d^\dagger s \right) ^{(2)}_\mu 
                    + \chi \left(d^\dagger \tilde d\right)^{(2)}_\mu~~,
\label{quadrB}
\end{equation}
and 
\begin{equation}
q^{(p)}_{\mu} = \left(p^\dagger \tilde p \right)^{(2)}_{\mu}~~.
\end{equation}
The operators $\tilde \gamma_m = (-1)^m \gamma _{-m}$ (where $\gamma$
stands for $s$, $d$ or $p$ bosons) are introduced so as to have
operators with the appropriate properties under spatial rotations.

The simplest form for the dipole operator is
\begin{equation}
D_{\mu} = \xi \left(p^{\dagger} + \tilde p\right)_\mu~~,
\end{equation}
where $\xi$ is an overall normalization  parameter, which will be
assumed equal to unity in the applications.

\subsection{The coupling in the laboratory frame}

First the laboratory IBM calculation is presented. 
The quadrupole $s-d$ hamiltonian for the O(6) limit can be written in the
form
\begin{equation}
H_{sd}[O(6)]= b ~ C_2[O(5)] + c ~ C_2[O(3)] + d ~ C_2[O(6)]
\label{HamO6}
\end{equation}
where $C_2$ stands for the quadratic Casimir operator of the indicated
algebra as defined in Ref. \cite{IA87}.
Energies associated with this $s-d$ hamiltonian
can be obtained analytically to be 
\begin{equation}
E= 2 b ~ \tau (\tau+3) + 2 c ~ J(J+1) + 2 d ~ \sigma (\sigma+4)~~,
\end{equation}
where $\sigma$, $\tau$ and $J$ are the labels of the irreducible
representations of the algebras O(6), O(5) and O(3)
respectively. The coupling to the dipole oscillations and the
calculation of transition probabilities is done
numerically by using the computer codes
GDR and GDRT, respectively \cite{Piet}. 

The states belonging to the $\sigma=N$ representation are 
favoured in energy and the relative position of the states belonging to
other representations is governed by the parameter $d$.  In the
O(6) limit these representations are not connected by the quadrupole
$sd$ operator if $\chi=0$ is taken\footnote{With this choice the
quadrupole operator is a generator of the O(6) algebra.} 
in Eq. (\ref{quadrB}), and they therefore remain unmixed 
(and characterized by a good value of $\sigma$) also with the inclusion of the 
dipole-quadrupole coupling. Consequently the dipole operator will
only connect
the ground state with states of the $\sigma=N$ representation, 
which will therefore be the only ones 
contributing to the $B(E1; 0_1^+ \rightarrow 1_n^-)$ distribution.
This implies that,
as far as the dipole distribution is concerned, the calculation
is completely insensitive to the value of the parameter $d$.
For the other parameters, in order to favour the comparison with 
preceding results obtained 
within the simple scheme based on the geometrical
model, the parameter $c$ has been put equal to zero in order
to minimize the splitting due to the angular momentum, a feature not
accounted  
for by the unprojected intrinsic state.
The results obtained for the  energies of dipole states and the
corresponding 
$B(E1; 0_1^+ \rightarrow 1_n^-)$ are shown in Fig. 3 (left panels) 
for different values
of the parameter $b$ in front of the O(5) Casimir operator.  
At variance with the previous distribution displayed in Fig. 2 we have
now a  
discrete distribution due to the finite number of bosons and consequently 
finite number of states. 
Note that the pure SU(3) limit with $\chi=-\sqrt{7}/2$ 
gives a dipole strength distribution
with only two lines, since only two $1^-$ states can be obtained by
coupling 
the dipole boson with the ground state rotational band. 
In the O(6) limit there are instead $(N+1)$ $1^-$ states 
connected to the ground state by the dipole operator, namely
all those originated by the $0^+$ and $2^+$ states appearing in the
$\sigma =N$ band. We can however compare the envelope of this discrete 
distribution with the continuous distribution given by the 
geometrical model
in its simplest adiabatic version. In order to facilitate comparisons
an appropriate discretization of Fig. 2 for the case of N=15 bosons, 
an unperturbed dipole energy $\epsilon_1$ = 15 MeV, 
and a dipole-quadrupole coupling $\kappa$=0.2 MeV (same parameters as
in Fig. 3) is presented in Fig. 4. 
In the limit of very small values of $b$, namely in 
strong-coupling situations,  
the splitting due to the different values of $\tau$ 
within each O(6) representation is small compared to the coupling. 
As a consequence the patterns obtained in Figs. 3 and 4 are
similar, since the mixing between the different states induced by the
quadrupole-dipole interaction is large.
At the other extreme, 
for values of $b$ large with respect to the dipole-quadrupole coupling
the pattern observed in Fig. 3 becomes very asymmetric, 
tending to concentrate the
transition strength onto the lowest state, washing out any resemblance
with the one 
obtained in the geometrical model, Fig. 4, and giving 
rise to a situation close to the 
pure spherical case. In fact in this weak-coupling limit, because of the
large $\tau$ splitting, only the lowest $1^-$
states have large overlaps with the initial s--d ground state.

For a better understanding of the situation we will study next the 
problem within the intrinsic frame in the IBM.

\subsection{The coupling in the intrinsic frame}

Let us now study the problem of the coupling of quadrupole and
dipole degrees of freedom in the intrinsic frame, 
within the IBM description. The basic idea of
the intrinsic-frame formalism in this case is to consider that the
pure quadrupole states of the ground ``band'' are 
globally described by a boson condensate of the form
\begin{equation}
|g \rangle = {1 \over \sqrt{N!}} (\Gamma_g^\dag) ^N |0 \rangle ~~,
\label{ground1}
\end{equation}
where the basic boson is given by
\begin{equation}
\Gamma_g^\dagger={1 \over \sqrt{1+ \beta^2}} 
\left[ s^\dagger + \beta \cos \gamma d^\dagger_0 + {1 \over \sqrt{2}} 
\beta \sin\gamma (d^\dagger_2 + d^\dagger_{-2})\right]~~,
\label{ground2}
\end{equation}
$\beta$ and $\gamma$ being obtained by minimizing the energy
surface.  In the case of SU(3) one has $\beta=\sqrt{2}$ and
$\gamma =0$, and the intrinsic state is fully representative of
the ground-state rotational band.  In the case of O(6), instead,
one has $\beta =1$, and the intrinsic state, which gives rise to a 
$\gamma$-independent energy surface, is
associated with all states of the ``band'' corresponding to
the $\sigma=N$ representation.
For the dipole part, the corresponding  building blocks 
are $p_0^\dagger$, $p_1^\dagger$ and
$p_{-1}^\dagger$. Thus, the three states $|g p_\mu \rangle$ 
provide with an intrinsic basis where the dipole-quadrupole states can
be studied.  In this
basis, the pure $sd$ and $p$ parts of the hamiltonian are diagonal and
the only non-diagonal contributions come from the interaction term.
The matrix elements needed in order $1/N$ are
\begin{equation}
\langle g p_\mu|Q^{(sd)} \cdot q^{(p)}|g p_\mu \rangle= 
\sum_{i=0,\pm 2} \langle g|Q^{(sd)}_i|g \rangle  
\langle p_\mu|q^{(p)}_{-i}|p_\mu \rangle~, 
\label{Qqdef}
\end{equation}
where 
$ \langle g|Q^{(sd)}_{i}|g \rangle\equiv Q^{(sd)}_{i}(\beta, \gamma)$ 
have already been calculated \cite{AL92},
\begin{equation}
Q^{(sd)}_{0}(\beta, \gamma)= {N \over 1+ \beta^2}\left[2 \beta \cos \gamma
                  - \sqrt{2 \over 7} \chi \beta^2 \cos 2\gamma\right]~,
\label{Qsd1}
\end{equation}
\begin{equation}
Q^{(sd)}_{2}(\beta, \gamma)= Q^{(sd)}_{-2}(\beta, \gamma)
= {N \over 1+ \beta^2}\left[\sqrt{2} 
\beta \sin \gamma  + \sqrt{1 \over 7} \chi \beta^2 \sin 2\gamma\right]~~.
\label{Qsd2}
\end{equation}
The $Q^{(sd)}$ matrix elements not specified are zero. The matrix
elements of $q^{(p)}$ can be easily computed to be
\begin{eqnarray}
\langle p_1|q^{(p)}_0|p_1 \rangle = \langle p_{-1}|q^{(p)}_0|p_{-1}
\rangle & = & \frac{1}{\sqrt{6}} ~, \\ 
\langle p_{0}|q^{(p)}_0|p_{0} \rangle & = & - \frac{2}{\sqrt{6}} ~,\\
\langle p_1|q^{(p)}_2|p_{-1} \rangle = \langle p_{-1}|q^{(p)}_{-2}|p_1
\rangle & = & 1 ~. 
\label{qp}
\end{eqnarray}
All the remaining relevant matrix elements for our study are zero.

The parameter $\chi$ is the structure constant in the quadrupole 
operator $Q^{(sd)}$. 
The SU(3) case (studied in Ref. \cite{SH87}) is 
obtained for $\chi= - {\sqrt{7} \over 2}$ and 
the O(6) limit corresponds to $\chi=0$. The dipole energies are
obtained by  diagonalising the interaction
$V_{sd-p}$ in the basis $|g p_\mu \rangle$. They are analytically 
given by
\begin{eqnarray}
\lambda_0 & = & \epsilon_1 + \kappa ~ \frac{2}{\sqrt{6}}~
Q^{(sd)}_{0}(\beta, \gamma) ~,
\nonumber \\
\lambda_+ & = & \epsilon_1 - \kappa ~\left[ \frac{1}{\sqrt{6}}~
Q^{(sd)}_{0}( \beta, \gamma)
+ Q^{(sd)}_{2}(\beta, \gamma)\right] ~,\label{enerp1} \\
\lambda_- & = & \epsilon_1 - \kappa ~\left[ \frac{1}{\sqrt{6}}~ Q^{(sd)}_{0}(\beta, \gamma)
- Q^{(sd)}_{2}(\beta, \gamma) \right] ~,\nonumber
\end{eqnarray}
which provide the dependence of the dipole energies on the IBM
deformation parameters $(\beta, \gamma)$.  With the choice 
$\chi= - {\sqrt{7} \over 2}$, corresponding to the SU(3) limit,
Eqs. (\ref{enerp1}) become equivalent to Eqs. (9b) in Ref. \cite{SH87}
with terms linear and quadratic in $\beta$. The corresponding energies
are shown in Fig. 5a. 
The O(6) limit corresponds instead to $\chi=0$. In this case
Eqs. (\ref{enerp1}) have only a linear term in $\beta$, and
the corresponding energies are shown in Fig. 5b.
These latter energies are equal to those obtained in the preceding
section for the geometrical model, once the correspondence is made
between deformation parameters and coupling strengths in the two models,
such that 

\begin{equation}
\Delta_{IBM} = 2 \sqrt{2 \over 3}~ {\beta_{IBM} \over 1 + 
\beta_{IBM}^2}\kappa N = \Delta   =  \epsilon_1 \sqrt{2 \over 15} {\kappa_1 
\over C_1} \beta_0 ~ .
\label{delta}
\end{equation}

As in the case of the geometrical model, a simple way of obtaining
the dipole strength distribution from the 
IBM intrinsic state is to average the intrinsic energies over
$\gamma$.  In analogy to the intrinsic energies, the resulting
distribution will show, {\it mutatis mutandis}, a pattern 
identical to the one shown in Fig. 2 for the geometrical model.

\subsection{Projection from the intrinsic frame to the laboratory}

We have seen so far that the straightforward
use of the intrinsic state in the IBM without any projection technique
gives results similar to the ones obtained in the geometrical model.
These results are consistent with those obtained in the laboratory
frame only for boson hamiltonians that give rise to splittings in the
elements of each O(6) multiplet which are much smaller than
the splitting due to the dipole-quadrupole coupling. In all other more
general cases, the effect of the breaking of degeneracy of each
multiplet is essential, and proper treatment of the projection to the
laboratory system from the intrinsic state is needed.    

In this subsection it is therefore shown how to project the intrinsic
state for the ground-state representation $\sigma=N$ into the $\tau$
component. With this we would like to show that the calculations in
the intrinsic frame, after projecting after on $\tau$, reproduce the
laboratory results. 
In the O(6) case the projector is known \cite{Bes}, and the important
feature is that it produces exactly the laboratory states 
$|\sigma=N, \tau, L \rangle$, belonging to the representation
$\sigma=N$. 

We have considered states of the form
$|\sigma=N, \tau, L \rangle$ coupled to $|p \rangle$ to total
$J=1$. The states 
$|\sigma=N, \tau, L \rangle$ are obtained by applying the projector
$\Phi_{\tau,L,M}(\gamma, \theta_i)$ (B\`es functions 
\cite{Bes}) to the intrinsic ground
state $|g \rangle$ as given in Eqs. (\ref{ground1},\ref{ground2}). For each value of
$\tau$ there is only one state (either $L=0$ or $L=2$) to be
considered. Thus, for a given $N$ there will be $N+1$ dipole states.

In order to diagonalize the Hamiltonian in the $J=1$ subspace, 
we have included diagonal (trivial) and non--diagonal
terms.  For the latter ones
we have used the results of the matrix elements of the interaction
$Q^{(sd)}.q^{(p)}$ in the intrinsic frame (\ref{Qqdef}-\ref{qp}) and
made the appropriate 
integrations on $\gamma$ and the Euler angles $\theta_i$ to go from
the intrinsic to the laboratory frame. The resulting matrix to be 
diagonalized in the lab connects states of the same $\tau$ (diagonal
terms) and states of $\tau$ with $\tau \pm 1$ as known.

We have done full calculations up to $N=20$ for the case $\chi=0$ 
and found simple expressions for the non-diagonal matrix elements 
that allow calculations for any value of $N$. These expressions are
(we will use the notation $|\tau_{(L_{sd})}, l_p; J M \rangle$ for the
states, with the selection rule  
$\tau^\prime _{final}$~=~$\tau_{initial} \pm 1$)

\begin{equation}
\langle \tau^\prime _{(2)},1; 1 M|Q^{(sd)}.q^{(p)}| \tau_{(2)}, 1 ;1 M
\rangle= - \sqrt{2 \over 3} {\beta \over 1 + \beta^2} N~~,
\end{equation}

\begin{equation}
\langle \tau^\prime _{(0)}, 1; 1 M|Q^{(sd)}.q^{(p)}| \tau_{(2)}, 1 ;1
 M \rangle=  -  2 {\beta \over 1 + \beta^2} N~
{(-1)^R \sqrt{R} \over \sqrt{2 \tau^\prime _{(0)} +3}} ~,
\end{equation}
where $R=(3 \tau_{(2)} + 3 - \tau^\prime _{(0)})/6 $.
We have checked that these expressions coincide, in leading order in
1/N, with those associated with the corresponding ``exact'' states for
$\tau= 0, 1$ and $2$. With these matrix elements we have done a
series of calculations and the resulting dipole 
$B(E1;0_1^+ \rightarrow 1_n^-)$ distributions are compared in Fig. 3
(right panels) with the exact results in the 
laboratory frame for different values of $b$.
In Fig. 4 the corresponding distribution obtained from the 
unprojected intrinsic
state (equal for all values of $b$) is plotted for comparison.
In all the cases we have fixed the number of bosons to 15 and the 
intensity of the quadrupole ($sd$)-quadrupole ($p$) interaction to
the value $k=0.2$ MeV.
We have changed the intensity of the $\tau (\tau +3)$
term by acting on the $b$ term in Eq. (\ref{HamO6}).  
In the following points we comment on the results given in Fig. 3.

\begin{itemize}
\item i) $b=0.00015$ MeV. This gives an excitation energy of the $2_1^+$
of 0.0012 MeV, almost degenerate case. The pattern  of the $B(E1)$ in
the exact calculation \cite{Piet} (left lowest panel) 
is as mentioned before 
(three maxima, two at the edges with $B(E1)$ about 3 times the 
value of the maximum at the center). In our
projected calculation (right lowest panel) 
we obtain similar pattern with three bumps but
now the maxima at the edges are lower than in the full lab calculation
and the maximum at the center is higher than the one in the exact 
calculation.
In this way we obtain three maxima of roughly the same height.
The two minima are more or less in the same places in both
calculations. It is remarkable that the number of states in the
projected calculation close to the edges is larger than in the full
lab calculation and the number of states close to the center is
smaller. In that way the summed $B(E1)$ values in each of 
the three bumps is the same in both calculations. The energy
separation between edges is also about the same in both calculations.

\item ii) $b=0.0015$ MeV. This gives an excitation energy of the $2_1^+$
of 0.012 MeV. Now the patterns are more similar in both calculations,
the population of the lowest part being favoured in both cases. 
The results start to deviate appreciably from the unprojected ones.

\item iii) $b=0.015$ MeV. This gives an excitation energy of the $2_1^+$
of 0.120 MeV. Now the patterns of the laboratory and the projected
intrinsic state are very close, with strong deviation from the pure
results of the intrinsic state.

\item iv) $b=0.15$ MeV. This gives an excitation energy of the $2_1^+$
of 1.2 MeV. Now only the 3-4 lowest states are populated and energies
and $B(E1)$'s are practically identical in both calculations.

\end{itemize}

From this comparison we conclude that our intrinsic calculation
followed by projection on $\tau$ gives a good approximation to the
laboratory results. It is worth noticing that in all cases the
agreement is excellent except for the almost degenerate case, which is
surprising since in that case the intrinsic state is expected to be a
good approximation (see Fig. 3 lowest left panel and Fig. 4). This 
can be due either to the projection method or to the intrinsic trial
wave function. In this case the projection method is known to be exact
for the large $N$ limit, thus the problem must come from the trial wave
function. As mentioned above, in the limit of very small values of $b$
(strong-coupling situations)  
the splitting of the different values of $\tau$ 
within each O(6) representation is small compared to the coupling. 
Consequently, the trial wave function proposed as $|g p_\mu \rangle$
with $|g \rangle$ given by Eqs. (\ref{ground1},\ref{ground2}) could
be not completely appropriate in this case.  
The coupling to the additional 
dipole boson has in fact destroyed the full $\gamma$-unstability of the system
(cf. Fig. 5), leading to an additional variation of the order of 1/$N$ in the 
basic boson of the condensate.  The situation is similar to that
obtained, for example, in the coupling of an odd particle to a SU(3)
boson core, where the additional particle slightly shiftes the position of the
minimum of the energy surface in the $\beta$-$\gamma$ plane. 

It should be noted that realistic
cases of O(6) nuclei correspond to values of $b$ around $0.05$
MeV, far from degeneration in $\tau$. We have checked this by making a
set of calculations for a fixed and reasonable value of $b$ and
changing the interaction $\kappa$. In those cases agreement between
laboratory calculations and intrinsic plus projection is fine for any 
reasonable value of $\kappa$. Only for extremely large and unphysical
values of $\kappa$ some discrepancies between both calculations start
to appear. 

Once the energy spectra and the dipole transition strengths have been
discussed we present in the next Section the cross sections for
absorption of unpolarized $\gamma$ radiation in the GDR region.

\section{Photonuclear cross-section}

Dipole strength distributions are traditionally measured in
photoabsorption processes. 
Although photoabsorption cross-sections directly reflect the
dipole strength, some of the features associated with this strength
may disappear, be masked or modified by the effect of the finite
widths. For this reason, once the distribution of the
individual dipole states has been analyzed in the preceding section,
we prefer in this section to compare results
obtained within the IBM dipole model in the laboratory frame,
unprojected  and projected intrinsic frame directly for 
photoabsorption cross-sections.  

The cross section for photoabsorption is in
fact given by \cite{sigma} 
\begin{equation}
\sigma(E) = {8 \pi e^2 \over 3 \hbar c}~\sum_n
~E_n~|\langle 1_n^-||D||0_1^+ \rangle|^2~
{\Gamma_n ~ E^2 \over (E^2-E_n^2)^2 + \Gamma_n^2~ E^2}~,
\end{equation}
where $\Gamma_n$ are the widths of the dipole states at energy $E_n$.
The dipole widths are either taken as a constant
or assumed to have a power law dependence  on the energy,
\begin{equation}
\Gamma_n = \Gamma_0 \left({E_n \over E_0}\right)^\delta~,
\end{equation}
where $E_0$ is the energy of the unperturbed dipole state and 
$\Gamma_0$ its width. Both $\Gamma_0$ and $\delta$ are used
as adjustable parameters when one is fitting experimental data.

In Fig. 6 we present the cross-sections obtained
for the two extreme ($b$=0.00015 MeV and $b$=0.15 MeV) 
dipole strength distributions  
corresponding to Fig. 3. Full lines give the
results for the case $b$=0.00015 MeV (almost $\tau$-degenerate case)
while dashed lines are for $b$=0.15 MeV.
In both cases $E_0=15$ MeV and
the intensity of the quadrupole (sd)-quadrupole (p) interaction
($\kappa$) is 0.2 MeV.
Left panels give the laboratory results and right panels the
intrinsic plus projection on $\tau$ results.  

We have done two sets of calculations which select different widths
for the dipole states. Upper panels correspond to
the case in which the dipole widths are assumed to have a power law
dependence on the energy. Since we are not working with an 
specific nucleus, we have taken
$\Gamma_0$= 0.026 $\times ~ E_0^\delta$ (MeV) and
$\delta$=1.91. These
values come from one global parametrization of experimental
photoabsorption cross sections for $A > 50$ \cite{photofit}.
That implies widths ranging from about 3.5 to 6.0 MeV for 
$E_n$ ranging from 13 to 17 MeV, respectively.
Lower panels represent equivalent cases but taking 
the dipole widths as a constant equal to 2.5 MeV. 

 First of all we should realize that the actual shape
of the photoabsorption cross-section is very much 
dependent on 
the widths assumed. The dashed lines (case with $b$=0.15 MeV) in all
panels still reflect contributions 
from the two lower states since they are separated by about
3 MeV. By contrast, all the states contribute
to the almost $\tau$ degenerate case (full lines, $b$=0.00015 MeV)
since now the states differ by less than 0.5 MeV.
Thus the image on the photoabsorption cross-section of the
shape of the dipole strength distribution in Fig. 3 is 
diluted, even showing a plateau when we use the biggest
widths (upper panels).

In all the cases, independently of the election of $b$ and
the dipole widths assumed (either constant or energy dependent),
a good agreement between the laboratory and $\tau$-projected intrinsic 
frame calculations is observed when comparing 
left and right panels, although as in Fig. 3 as the absolute 
value of the ratio b/$\kappa$ decreases its quality becomes a bit
worse. It is worth noticing that the calculations for the case of
unprojected geometrical adiabatic situation given in Fig. 4 are
indistinguishable from those plotted with full line on the 
right-hand-side panels ($b$=0.00015 MeV) as expected.

\section{Summary and conclusions}

We have studied the problem of the coupling of the quadrupole and
dipole oscillations in $\gamma$-unstable nuclei in the framework of
the IBM. We have obtained and explained the results in the laboratory by
studying the problem in the intrinsic frame taking at higher order the
non-degeneracy in $\tau$.

In the $\tau$-degenerate limit the dipole mode total splitting
(2$\Delta$) turns out to be proportional to the dipole-quadrupole
coupling strength, the number of sd bosons considered, N,  and the
quadrupole deformation. Furthermore, the dipole mode splits in N+1
lines and its distribution is symmetric around the unperturbed dipole
energy. This symmetry is destroyed as soon as we allow non-degeneracy
in $\tau$, giving rise to a very asymmetric distribution. In the weak
coupling limit the transition strength is concentrated onto the lowest
states. Since this limit seems to correspond 
to the actual situation in typical O(6)
nuclei, $^{196}$Pt or $^{134}$Ba for instance, 
the explicit treatment of the non-degeneracy in $\tau$ is crucial.
In this case the results from the geometrical model 
in the adiabatic limit are quite different from the dipole boson model ones. 
In this sense, the O(6) limit is different to the SU(3) case in which the 
rotational excitation energy is low and can be ignored in first 
approximation and, consequently, the geometrical model in the adiabatic 
limit gives an appropriate description. 

Calculations of photoabsorption cross-sections using the intrinsic
state in the IBM followed by projection on $\tau$ give a good
approximation to the laboratory results. No experimental data on O(6)
nuclei are available and they could be valuable to elucidate whether the
dipole strength distribution in actual O(6) nuclei is in agreement
with the $\tau$-projected dipole boson model results presented here.

\section*{Acknowledgments}

This work was supported in part by a Spanish--Italian CICYT-INFN 
agreement and 
by the Spanish DGICYT under project number PB98-1111. 
Part of this work was done while E.G.L. was at Sevilla University as a
Marie Curie Fellow with contract ERBFMBICT-983090.
We acknowledge useful discussions with F.\ Iachello and A. Leviatan.
We thank P. Van Isacker for providing us with the IBM codes GDR and GDRT.

\begin{figure}[]
\begin{center}
\mbox{\epsfig{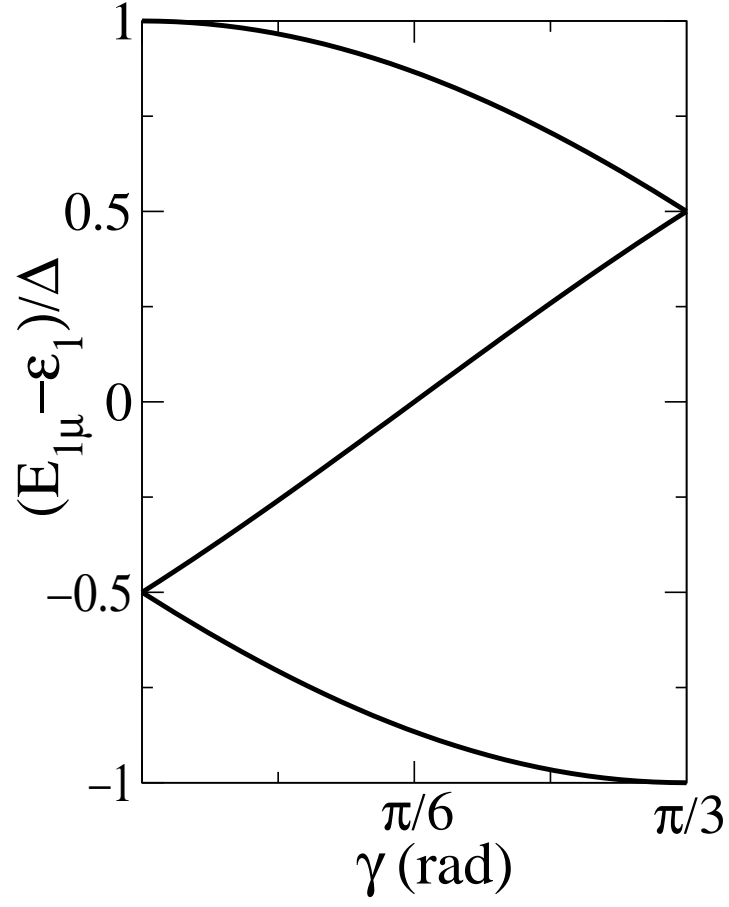}}
\end{center}
\caption{Energies of the dipole resonances for an ellipsoidal nucleus
with quadrupole deformation $\beta_0$ as a 
function of the shape parameter $\gamma$ in the geometrical model.  Energies
are given with respect to the unperturbed energy $\epsilon_1$ in units
of $\Delta$ $(\sqrt {2 \over 15} \epsilon_1~\beta_0~\kappa_1/C_1)$. }
\label{fig1}
\end{figure}

\newpage

\begin{figure}[]
\begin{center}
\mbox{\epsfig{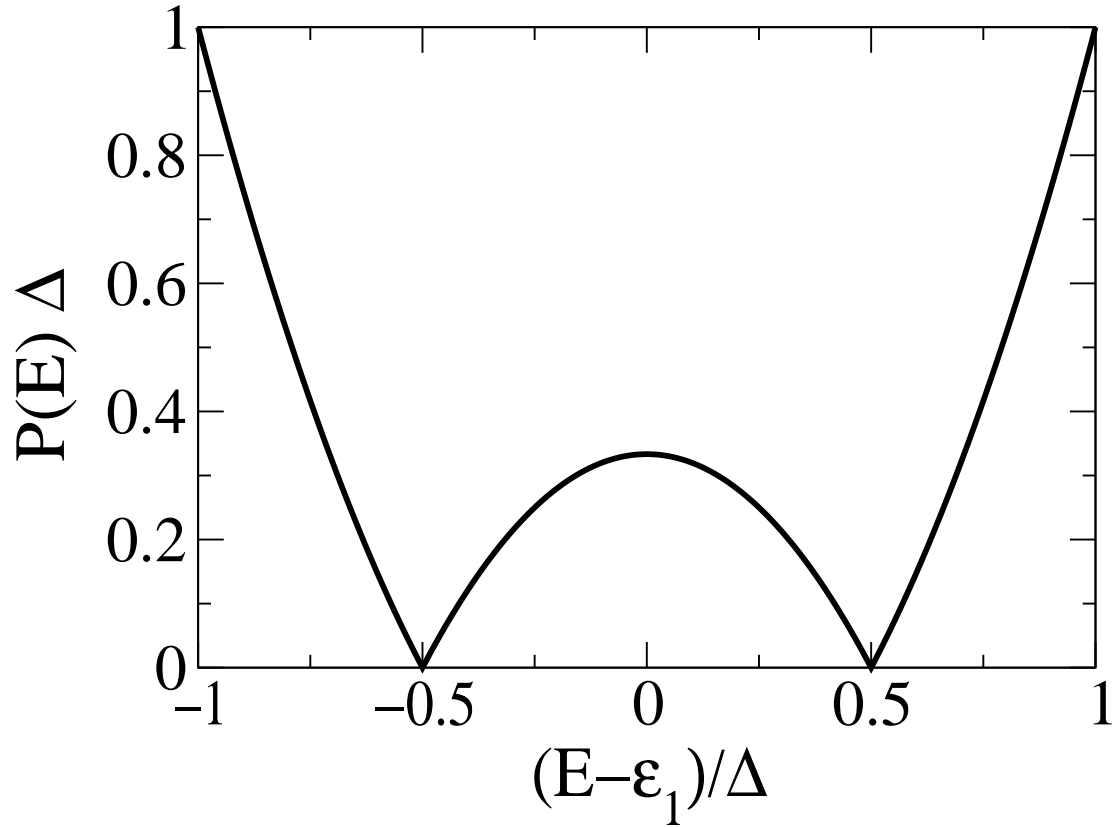}}
\end{center}
\caption{Probability per unit energy for absorption of a quantum E, in
units of 1/$\Delta$, as
a function of the energy in the geometrical model for a $\gamma$-unstable
nucleus.  Energies
are given with respect to the unperturbed energy $\epsilon_1$ in units
of $\Delta$.  }
\label{fig2}
\end{figure}

\newpage

\begin{figure}[]
\begin{center}
\mbox{\epsfig{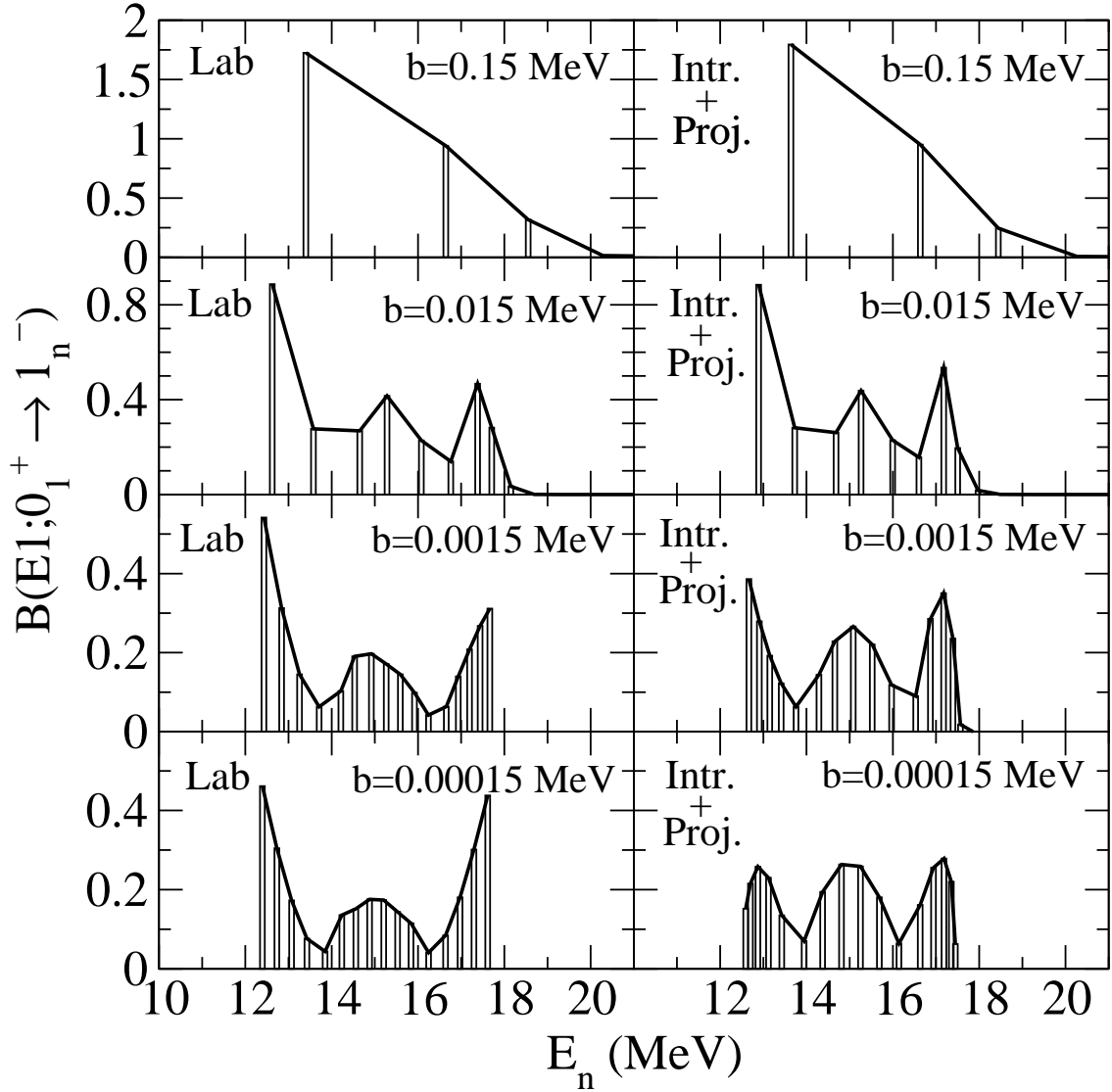}}
\end{center}
\caption{Distribution of B(E1) strength obtained in the IBM dipole-boson 
model.  The calculations have been performed in the case of N= 15 bosons, 
with an O(6) hamiltonian with parameters c=0 and
different values of b, an unperturbed dipole energy $\epsilon_1$ = 15 MeV 
and a dipole-quadrupole coupling $\kappa$= 0.2 MeV. Left panels
present the exact calculation in the laboratory system and right ones
the corresponding calculations in the intrinsic frame plus
projection on $\tau$. The continuous lines just joint the extremes of
the bars.}
\label{fig3}
\end{figure}

\newpage

\begin{figure}[]
\begin{center}
\mbox{\epsfig{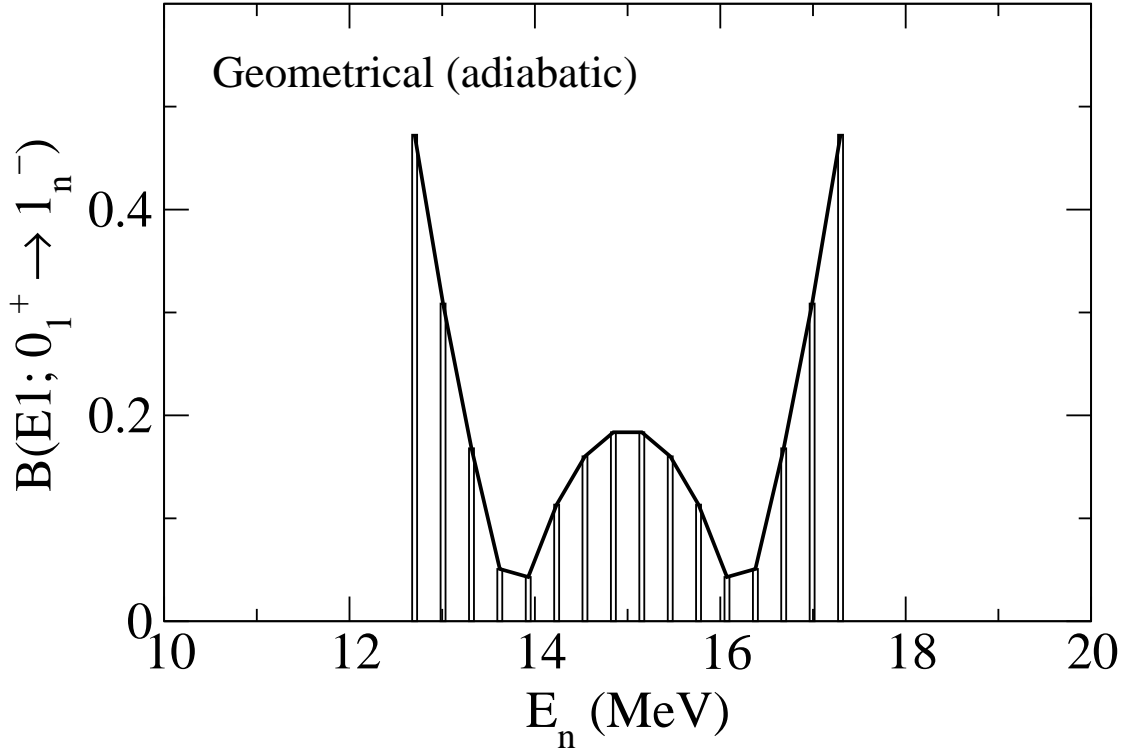}}
\end{center}
\caption{Distribution of B(E1) strength obtained in the 
adiabatic picture of the geometrical model or, equivalently, in the
unprojected IBM dipole-boson model. 
The bar diagram is obtained by discretizing the
continuous distribution of Fig. 2 into 16 states (appropriate to the
case N=15 presented in Fig. 3) and summing the contribution of each 
discrete state in each bin (full line just joint the extremes 
of the bars). The calculations have been performed for the case of 
unperturbed dipole energy $\epsilon_1$ = 15 MeV 
and dipole-quadrupole coupling $\kappa$=0.2 MeV (same parameters as
the ones in Fig.3). The relation between $\kappa$ and 
${\kappa_1 \over C_1} \beta_0$ is given in Eq. (\ref{delta}).} 
\label{fig4}
\end{figure}

\newpage

\begin{figure}[]
\begin{center}
\mbox{\epsfig{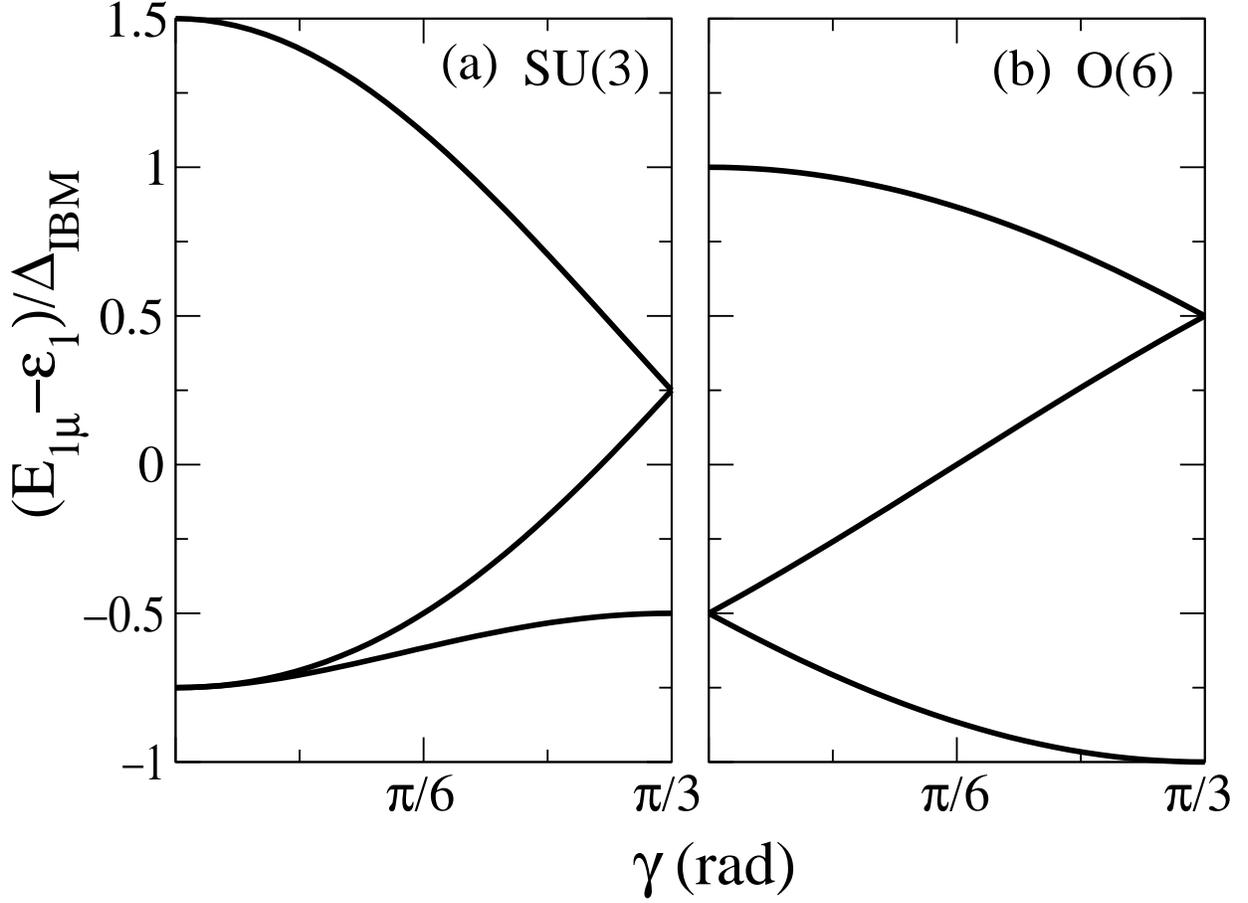}}
\end{center}
\caption{Energies of the dipole resonances 
obtained in the intrinsic frame from an IBM dipole-boson model
as a function of the shape shape parameter $\gamma$.  
Fig. (a) refers to the case of SU(3) and Fig. (b) to the case of O(6). Energies
are given with respect to the unperturbed energy $\epsilon_1$ in units
of $\Delta_{IBM}$  }
\label{fig5}
\end{figure}

\begin{figure}[]
\begin{center}
\mbox{\epsfig{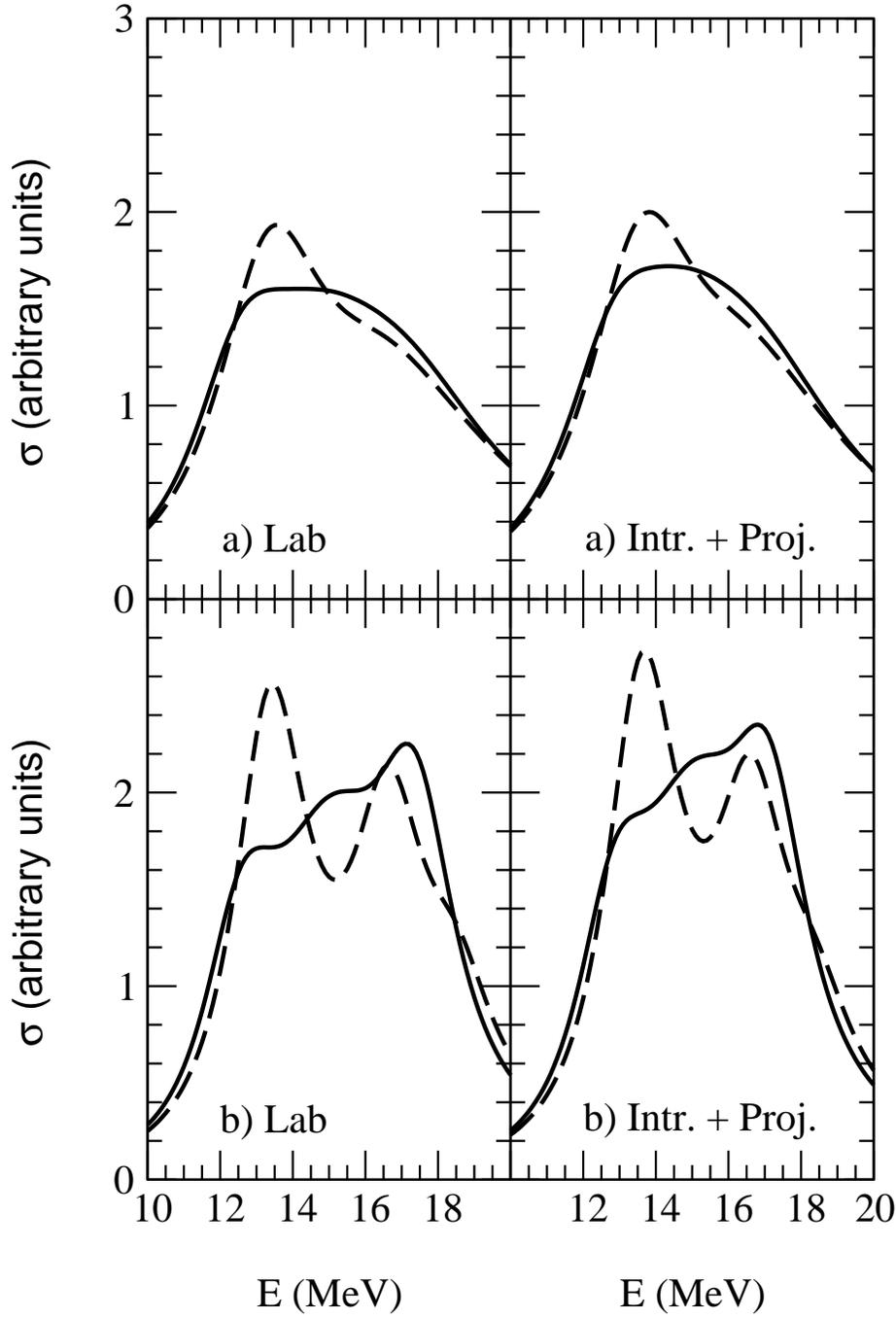}}
\end{center}
\caption{Photonuclear cross-sections for two selected cases from
Fig. 3. Left panels give the laboratory results and right panels the
intrinsic plus projection on $\tau$ results. Upper panels correspond to
the case in which the dipole widths are assumed to have a power
dependence law on the energy with 
$\Gamma_n$=0.026 $E_n^{1.91}$ (MeV). Lower panels 
represent equivalent cases but taking 
the dipole widths as a constant equal to 2.5 MeV. Full lines give the
result for the case $b$=0.00015 MeV (almost $\tau$ degenerate case)
while dashed lines are for $b$=0.15 MeV.}
\label{fig6}
\end{figure}


\bigskip

\begin{references}

\bibitem{BMII}
{A.\ Bohr and B.\ Mottelson, {\it Nuclear Structure, vol II},
Benjamin, Reading, Mass. 1975}

\bibitem{IA87}
{F.\ Iachello and A.\ Arima, {\it The interacting boson model},
Cambridge University Press, Cambridge, England, 1987}

\bibitem{WJ56}
{L.\ Wilets and M.\ Jean, Phys.\ Rev.\ {\bf 102} (1956) 788.}

\bibitem{AIu5}
{A.\ Arima and F. Iachello, Ann.\ Phys.\ (N.Y.) {\bf 99} (1976) 253.}

\bibitem{AIsu3}
{A.\ Arima and F. Iachello, Ann.\ Phys.\ (N.Y.) {\bf 111} (1978) 201.}

\bibitem{AIo6}
{A.\ Arima and F. Iachello, Ann.\ Phys.\ (N.Y.) {\bf 123} (1979) 468.}

\bibitem{GK80}
{J.\ N.\ Ginocchio and M.\ W.\ Kirson, Nucl.\ Phys.\ {\bf A 350} 
(1980) 31.}

\bibitem{DSI80} 
{A.\ E.\ L.\ Dieperink, O.\ Scholten and F.\ Iachello, Phys.\ Rev.\
 Lett.\ {\bf 44} (1980) 1747.}

\bibitem{BM80} 
{A.\ Bohr and B.\ Mottelson, Phys.\ Scripta {\bf 22} (1980)  468. }

\bibitem{DG64}
{M. Danos and W. Greiner, Phys.\ Lett.\ {\bf 8} (1964) 113.}

\bibitem{DG64b}
{M. Danos and W. Greiner, Phys.\ Rev.\ {\bf B 134} (1964) 284.}

\bibitem{MW82}
{I.\ Morrison and J.\ Weise, J. of Phys.\ {\bf G 8} (1982) 687.}

\bibitem{SH83}
{F.\ G.\ Scholtz and F.\ J.\ W.\ Hahne, 
Phys.\ Lett.\ {\bf B 123} (1983) 147.}

\bibitem{Mai84}
{G.\ Maino, A.\ Ventura, L.\ Zuffi and F. Iachello, 
Phys. Rev. {\bf C 30} (1984) 2101.}

\bibitem{SH87}
{F.\ G.\ Scholtz and F.\ J.\ W.\ Hahne, 
Nucl.\ Phys.\ {\bf A 471} (1987) 545.}

\bibitem{MGO94}
{M.\ Sugita, A. Gelberg and T. Otsuka,  
Nucl.\ Phys.\ {\bf A 567} (1994) 33.}

\bibitem{BrOt97}
{N. Yoshida, A. Gelberg,  T. Otsuka, I. Wiendenh\"over, H. Sagawa and
P. von Brentano, 
Nucl.\ Phys.\ {\bf A 619} (1997) 65.}

\bibitem{Piet}
{P.\ Van Isacker, Fortran Computer Codes GDR and GDRT,
unpublished.}

\bibitem{AL92}
{C.\ E.\ Alonso, J.\ M.\ Arias, F.\ Iachello and A.\ Vitturi, 
Nucl.\ Phys.\ {\bf A 539} (1992) 59.}

\bibitem{Bes}
{D.\ B\`es, Nucl.\ Phys.\ {\bf 10} (1959) 373.}

\bibitem{sigma}
{V.\ Redwani, G.\ Gneuss and H.\ Arenh\"ovel, Nucl.\ Phys.\ 
{\bf A 180} (1972) 254.} 

\bibitem{photofit}
{Reference Input Parameter Library for Theoretical Calculation of
Nuclear Reactions (RIPL Handbook), Chapter 6: Gamma-Ray Strength
Functions. \\ {\it http://www-nds.iaea.or.at/ripl/} }
 
\end{references}
\end{document}